\documentclass{nature_mod}

\usepackage{dcolumn}
\usepackage{bm}
\usepackage{url}
\usepackage{color}
\usepackage{lineno}
\usepackage[T1]{fontenc}
\usepackage{lineno}
\usepackage{epsfig}
\usepackage{acronym}
\usepackage{amsmath}
\usepackage{cuted}
\usepackage{mathtools}

\newcommand{\geant}{\ensuremath{\textsc{Geant4}}}

\newcommand{\tfig}{Figure~}

\newcommand{\degree}{\ensuremath{^{\circ}}}
\newcommand{\inchsign}{^{\prime\prime}}

\title{Electromagnetic character of the competitive $\gamma\gamma/\gamma$-decay from $^{137\mathrm{m}}$Ba}
\author{P.-A.~S\"{o}derstr\"{o}m$^{1\ast}$, L.~Capponi$^{1}$, E.~A\c{c}{\i}ks\"{o}z$^{1}$, T.~Otsuka$^{2,3,4}$, N.~Tsoneva$^{1}$, Y.~Tsunoda$^{2}$, D.~L.~Balabanski$^{1}$, N.~Pietralla$^{5}$, G.~L.~Guardo$^{1,6}$, D.~Lattuada$^{1,6,7}$, H.~Lenske$^{8}$, C.~Matei$^{1}$, D.~Nichita$^{1,9}$, A.~Pappalardo$^{1}$ \& T.~Petruse$^{1,9}$}

\begin{document}
\onecolumn
 \spacing{1}
\maketitle

\begin{affiliations}
  \item Extreme Light Infrastructure-Nuclear Physics (ELI-NP)/Horia Hulubei National Institute for Physics and Nuclear Engineering (IFIN-HH), Str. Reactorului 30, 077125 Bucharest-M\u{a}gurele, Romania
  \item Center for Nuclear Study, University of Tokyo, Hongo, Bunkyo-ku, Tokyo 113-0033, Japan
  \item Department of Physics, University of Tokyo, Hongo, Bunkyo-ku, Tokyo 113-0033, Japan
  \item RIKEN Nishina Center, 2-1 Hirosawa, Wako, Saitama 351-0198, Japan
  \item Institut f\"{u}r Kernphysik, TU Darmstadt, 64289 Darmstadt, Germany
  \item Istituto Nazionale di Fisica Nucleare, Laboratori Nazionali del Sud, 95125 Catania, Italy
  \item Universit\'{a} degli Studi di Enna KORE, Viale delle Olimpiadi, 94100 Enna, Italy
  \item Institut f\"ur Theoretische Physik, Universit\"at Gie\ss en, 35392 Gie\ss en, Germany
  \item Politehnica University of Bucharest, Splaiul Independentei 313, 060042 Bucharest, Romania
\end{affiliations}



\section*{Abstract}

\begin{abstract}
Second-order processes in physics is a research topic focusing attention from several fields worldwide including, for example, non-linear quantum electrodynamics with high-power lasers, neutrinoless double-$\beta$ decay, and stimulated atomic two-photon transitions. For the electromagnetic nuclear interaction, the observation of the competitive double-$\gamma$ decay from $^{137\mathrm{m}}$Ba has opened up the nuclear structure field for detailed investigation of second-order processes through the manifestation of off-diagonal nuclear polarizability. 
Here we confirm this observation with an $8.7\sigma$ significance, and an improved value on the double-photon versus single-photon branching ratio as $2.62\times10^{-6}(30)$.
Our results, however, contradict the conclusions from the original experiment, where the decay was interpreted to be dominated by a quadrupole-quadrupole component. Here, we find a substantial enhancement in the energy distribution consistent with a dominating octupole-dipole character and a rather small quadrupole-quadrupole element in the decay, hindered due to an evolution of the internal nuclear structure. The implied strongly hindered double-photon branching in $^{137\mathrm{m}}$Ba opens up the possibility of the double-photon branching as a feasible tool for nuclear-structure studies on off-diagonal polarizability in nuclei where this hindrance is not present.
\end{abstract}

\section{Introduction}

Polarizability is a fundamental concept in physics and chemistry defined from the principles of electromagnetic interaction.  It describes how applied electric or magnetic fields induce an electric or magnetic dipole, or higher-order multipole, moment in the matter under investigation\cite{jackson_electrodynamics}. 
In nuclear physics, the simple concept of polarizability influences observables over a broad range of topics. For example, the static dipole polarisation of the shape of the ground and excited states in atomic nuclei is influenced by the coupling to high-energy collective modes like the \ac{GDR} via virtual excitations.
In this case the nuclear static dipole polarizability, $\alpha_{\mathrm{d}}$, is obtained\cite{birkhan48ca} from the photonuclear population of excited states,
\begin{equation}
 \alpha_{\mathrm{d};\mathrm{E1}} = 2e \sum_{n} \frac{\left|\langle I_{0} \| \mathrm{E1} \| I_{n} \rangle \right|^{2}}{E_{n}-E_{0}},\label{eq:static_dipole}
\end{equation}
where the transition matrix elements of the wave functions correspond to the electric dipole transition, $\mathrm{E1}$, between the ground state, $I_{0}$, and an excited state, $I_{n}$, with $e$ the elementary unit charge and $E_{n}$ the energy of the state.

By expanding the concept of polarizability beyond the scalar case, one can divide the polarizability tensor into separate components. Typically, these are either spatial components like the birefringence properties of crystals or electric and magnetic multipole components. 
Within the nuclear structure framework, this type of off-diagonal polarizabilities can appear in very weak second order processes. In the electromagnetic case, the off-diagonal nuclear polarizability can be defined analogous to equation (1) in terms of either electric and magnetic components, or components of different multipolarities as
\begin{equation}
 \alpha_{\mathrm{M2E2}} = \sum_{n} \frac{\langle I_{\mathrm{f}} \| \mathrm{E2} \| I_{n} \rangle \langle I_{n} \| \mathrm{M2} \| I_{\mathrm{i}} \rangle}{E_{n}-\omega}\label{eq:off_d_pol}
\end{equation}
or,
\begin{equation}
 \alpha_{\mathrm{E3M1}} = \sum_{n} \frac{\langle I_{\mathrm{f}} \| \mathrm{M1} \| I_{n} \rangle \langle I_{n} \| \mathrm{E3} \| I_{\mathrm{i}} \rangle}{E_{n}-\omega}\label{eq:off_d_pol2}.
\end{equation}
Due to the parity conserving properties of the strong force, these decays can only be observed between two different states, $I_{\mathrm{i}}$ and $I_{\mathrm{f}}$. In the definition above, the denominator depends on the interference frequency, $\omega$, of the emitted $\gamma$ rays and is assumed to be half of the initial state energy.
This type of second-order electromagnetic processes of atoms was discussed in the doctorate dissertation of Maria G\"{o}ppert-Meyer\cite{maria_gm_double_gamma_phd} where she estimated a probability for an atomic two-photon absorption process relative to the single-photon process to be approximately $10^{-7}$, later to be confirmed with the observation of this effect in CaF$_{2}$:Eu$^{2+}$ crystals\cite{two_photon_excitation_observation}.

\begin{figure*}
\begin{center}
 \epsfig{file=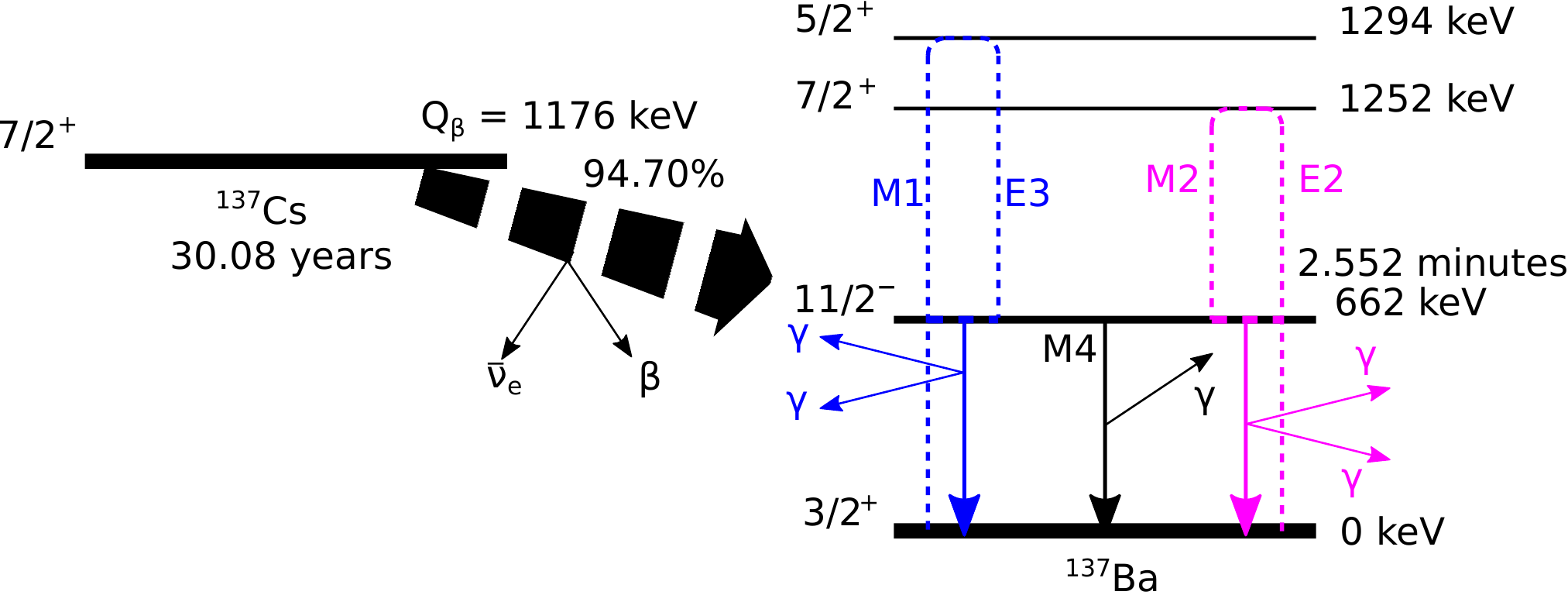,width=0.8\textwidth}\\
\end{center}
 \caption{{\bf{Decay diagram from the $^{137}$Cs ground state to the $^{137}$Ba ground state.}} Illustration of the single-$\gamma$ and the two types of double-$\gamma$ decay, as fed by the $\beta$ decay of $^{137}$Cs, including half lives of $^{137}$Cs and $^{137\mathrm{m}}$Ba. The energy of the $^{137}$Cs ground state ($Q_{\beta}$) is given relative to the $^{137}$Ba ground state. Here, M4 corresponds to the single-photon decay. The blue and pink decays show the lowest octupole-dipole and quadrupole-quadrupole compoenents, respectively. \label{fig:diagram}}
\end{figure*}

For many years, double-$\gamma$ decay was only observed in exceptional cases where both the ground state and the initial state have a spin-parity $J^{\pi}=0^{+}$ character for the doubly magic nuclei $^{16}$O \cite{16o_1,16o_2}, $^{40}$Ca \cite{40ca_90zr}, and $^{90}$Zr \cite{40ca_90zr}. Here single $\gamma$-emission is blocked, and only conversion-electron decay and double-$\gamma$ decay are allowed. In these experiments, the obtained information consists of correlations between energies and angles of these $\gamma$-rays, used to determine the decay probabilities of electric and magnetic dipoles. For a generalization of this phenomenon and the possibility to use it as a spectroscopic tool for more fundamental understanding of the underlying physics, large state-of-the-art \ac{HPGe} detector systems\cite{E5_decay,gammasphere_fail} have been used to search for the {\it{competitive}} $\gamma\gamma/\gamma$ decay where also the single $\gamma$ decay is allowed. Even though unsuccessful in that respect, these experiments successfully measured an E5 transition with the branching of $1.12(9)\times10^{-7}$. It is only with instrumentation developments of detector materials that can provide both the energy and time resolution required\cite{cebr} that the observation of the $\gamma\gamma/\gamma$ decay mode was announced \cite{walz_nature}. The setup used for that experiment consisted of five LaBr$_{3}$:Ce detectors arranged in a planar configuration with relative angles of 72\degree\ between the detectors, providing angular distribution data points at 72\degree\ and 144\degree. Thus, the collaboration could announce a $\gamma\gamma/\gamma$ decay signal with 5.5~$\sigma$ (standard deviations) statistical significance, near but above the typical discovery limit of 5~$\sigma$. From the two angular data points as well as the energy spectrum of the individual $\gamma$ rays at 72\degree\ angle, the off-diagonal polarizabilities $\alpha_{\mathrm{M2E2}}=33.9(2.8)$~$e^{2}$fm$^{4}$/MeV and the $\alpha_{\mathrm{E3M1}}=10.1(4.2)$~$e^{2}$fm$^{4}$/MeV polarizabilities were extracted. While the observation of the peak associated with $\gamma\gamma/\gamma$ decay was statistically clear, the nature of this decay was more uncertain, having the two dominating multipolarity combinations separated only by a small statistical difference, favouring the $\alpha_{\mathrm{M2E2}}$ component\cite{walz_thesis_book}. The decay diagram of this process is shown in Figure~\ref{fig:diagram}.

Given the nature of this experiment to observe a longstanding prediction of a the fundamental concepts in quantum mechanics and quantum electro-dynamics, and the possibility to extract nuclear structure observables from this, it is highly desirable to independently confirm this observation. Some possibilities that have been under discussion to perform this independent confirmation is to either return to the \ac{HPGe} approach with complex detector systems and event processing like the \ac{AGATA} setup \cite{ggg_agata_eunpc,agata} or highly charged radioactive ions \cite{triumf_ggg}. Here we report on an experiment using the \ac{ELIGANT} detector system\cite{gant_tdr,gant_matt} at the \ac{ELI-NP} facilities\cite{eli1,eli2,eli3} in a configuration\cite{eligant-gg} similar to what was used in reference~\cite{walz_nature}.
The experimental setup was optimised for obtaining a clean signal over a wide angular range\cite{eligant-gg} based on the expected intensities of the decay mode. Here, we can confirm the existence of the competitive double-photon decay process in atomic nuclei with an $8.7\sigma$ significance. We, however, find a significant octupole-dipole, E3M1, matrix element product contribution to the double-$\gamma$ decay mode of $\mathrm{^{137}Ba}$, contradicting the conclusions of the original experiment \cite{walz_nature}.
From our calculations using the \ac{EDF}+\ac{QPM} and the \ac{MCSM}, we find that both models reproduce the octupole-dipole component consistently, but the nature and the strength of the quadrupole-quadrupole component, differ significantly. 
It is interesting to note that this additional hindrance suggests a reduction of the $\gamma\gamma/\gamma$ branching with almost an order of magnitude in the most extreme case of Table~\ref{tab:alphas}, which is also the case that best reproduce the $\alpha_{\mathrm{E3M1}}$ polarizability. This opens for the possibility of a significant increase of the $\gamma\gamma/\gamma$ branching in nuclei in this region that do not exhibit this hindrance. In this case, experiments would be feasible also with more exotic sources \cite{PhysRevC.43.2586}, or even in-beam experiments within reasonable beam times, to follow the evolution of the quadrupole-quadrupole strength.

\begin{table*}[h!]
\centering
\begin{tabular}{|p{\textwidth}|}
\hline
\caption{\bf{Experimental and calculated $\alpha$ coefficients and $\gamma\gamma/\gamma$ decay branching ratios. \label{tab:alphas}}}
\hrule
 \begin{tabular*}{\textwidth}{@{\extracolsep{\fill}}lcccccccc}
& $B(\mathrm{M4})$& $\Gamma_{\gamma\gamma}^{\mathrm{exp}}/\Gamma_{\gamma}^{\mathrm{exp}}$ & $\Gamma_{\gamma\gamma}^{\mathrm{th}}/\Gamma_{\gamma}^{\mathrm{th}}$ & $\Gamma_{\gamma\gamma}^{\mathrm{th}}/\Gamma_{\gamma}^{\mathrm{exp}}$ &$\alpha_{\mathrm{M2E2}}$  & $\alpha_{\mathrm{E3M1}}$\\
& ($10^{3}$ e$^{2}$fm$^{4}$) & ($10^{-6}$) & ($10^{-6}$) & ($10^{-6}$) & (e$^{2}$fm$^{4}$/MeV) & (e$^{2}$fm$^{4}$/MeV)\\
\hline
This work &  & 2.62(30) & &  & $\pm8.8(50)$ & $\pm36.4(20)$\\
EDF+QPM (0.6$g_{\mathrm{s}}^{\mathrm{bare}}$) & 1.15 & & {\bf{3.73}} & 5.13 & {\bf{59.4}} & 20.7\\
EDF+QPM ($g_{\mathrm{s}}^{\mathrm{bare}}$) & 3.30 & & 1.34 & 15.2 & 104 & {\bf{32.8}}\\
MCSM (0.6$g_{\mathrm{s}}^{\mathrm{bare}}$) & 1.18 & & 0.579 & 0.840 & -2.14 & -21.2\\
MCSM ($g_{\mathrm{s}}^{\mathrm{bare}}$) & 3.28 & & 0.196 & {\bf{2.20}} & {\bf{-3.34}} & {\bf{-34.3}}\\
Literature\cite{walz_nature} & 0.98 & 2.05(37) & &  & 33.9(28) & 10.1(42)\\
QPM\cite{walz_nature} & 1.11 & & 2.69 & & 42.6 & 9.5\\
\hline
 \end{tabular*}
\footnotesize{
The $\Gamma_{\gamma\gamma}/\Gamma_{\gamma}$ decay branching ratio is shown both with unquenched ($g_{\mathrm{s}}^{\mathrm{eff}}=g_\mathrm{s}^{\mathrm{bare}}$) and quenched gyromagnetic spin factors ($g_{\mathrm{s}}^{\mathrm{eff}}=0.6g_\mathrm{s}^{\mathrm{bare}}$). The latter limit was chosen based on the reproduction of individual reduced transition probabilities. Depending on the calculation the values of $g_{\mathrm{s}}^{\mathrm{eff}}$ to best reproduce nuclear data are typically within this range. Thus, these limits should be representative of the uncertainties in the theoretical calculations, giving a range of $\sim50$\% for both the $\alpha_{\mathrm{M2E2}}$ and $\alpha_{\mathrm{E3M1}}$ values for both models between the two extremes. The listed values closest to the measured branching are shown in {\bf{bold}} font. The best fit for the decay branching ratio for the EDF+QPM calculations, not listed here, is obtained when choosing $g_{\mathrm{s}}^{\mathrm{eff}}=0.7g_\mathrm{s}^{\mathrm{bare}}$ as $\Gamma_{\gamma\gamma}/\Gamma_{\gamma}^{\mathrm{th}}= 2.8$.
}
\hrule
\end{tabular}
\end{table*}

\section{Results}

\subsection{Experimental setup.} 
\begin{figure}[h!t]
\begin{center}
\epsfig{file=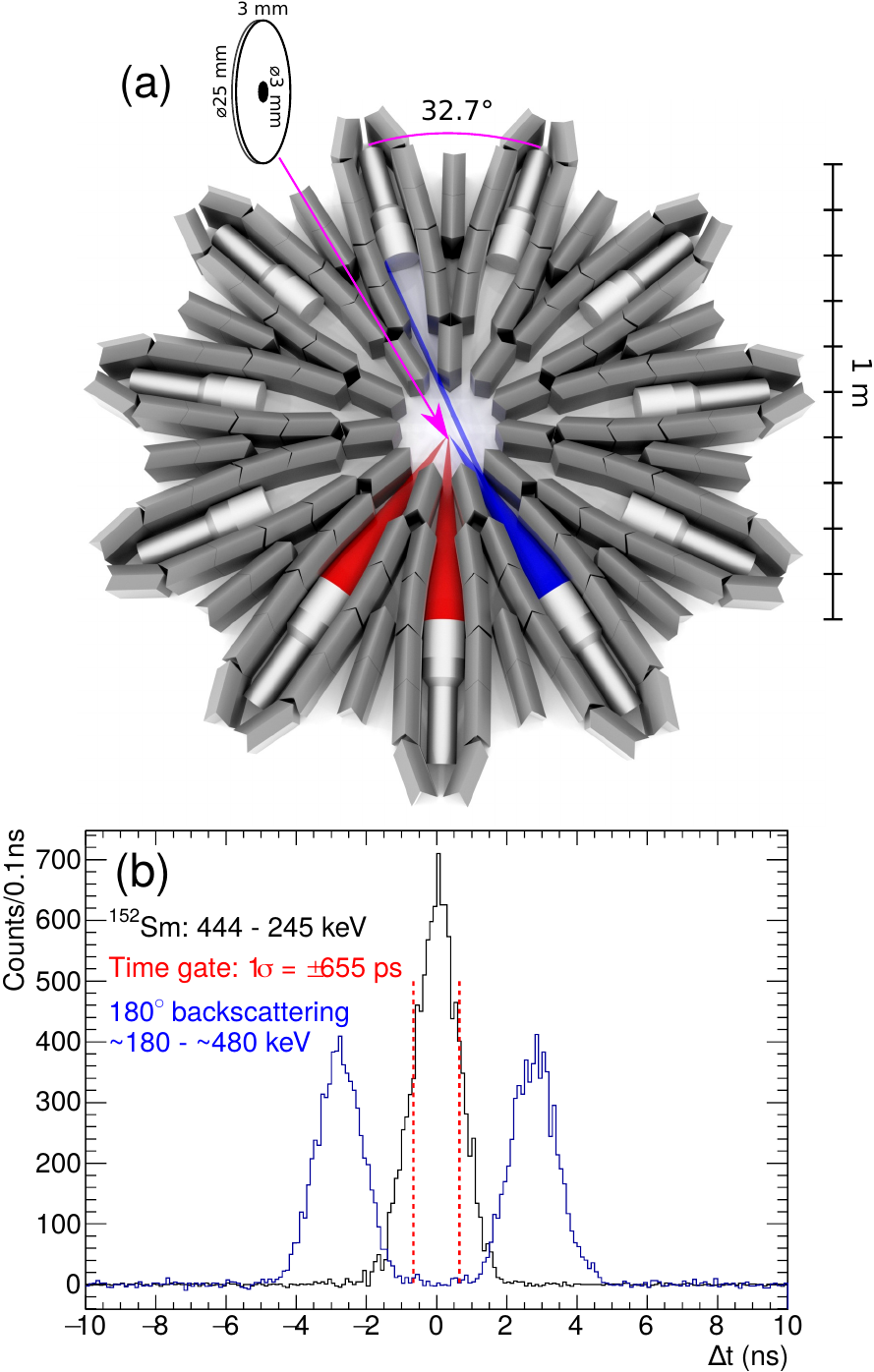,width=0.5\textwidth}
\end{center}
\caption{{\bf{Experimental setup.}} \bf{(a)} Coincident $\gamma$ rays could originate either from true double-$\gamma$ decay events illustrated with red cones, or from multiple Compton scattering between detectors illustrated with blue cones. \bf{(b)} Multiple Compton scattering events were rejected by the time difference ($\Delta t$) between the $\gamma$-ray interactions, shown in the blue histogram. The time condition for prompt $\gamma$-rays are shown as red dashed lines and verified with a $^{152}$Eu source.\label{fig:setup}}
\end{figure}

The experiment was performed using eleven $3\inchsign\times3\inchsign$ CeBr$_{3}$ detectors from \ac{ELIGANT}, shown in Figure~\ref{fig:setup}a. While \ac{ELIGANT} consists of both LaBr$_{3}$:Ce and CeBr$_{3}$ detectors, the CeBr$_{3}$ detectors were chosen to remove any possible source of background contribution from the natural radioactivity in lanthanum. The detector configuration was a circle with an inner radius to the front-face of the scintillators of 40~cm. This distance was enough to separate true coincidences from multiple Compton scattering of single $\gamma$ rays using the photon \ac{TOF}, see Figure~\ref{fig:setup}b. The relative angles between the eleven detectors were 32.7\degree, with an opening angle, given by the lead shielding, of $\pm3.4$\degree. This gave five independent $\gamma\gamma$-correlation angles centered at: 32.7\degree, 65.5\degree, 98.2\degree, 130.9\degree, and 163.6\degree. The detectors were separated with a minimum of approximately 15~cm of effective lead shielding between two neighbouring detectors to remove any contribution from single Compton scattering between detector pairs at low angles. The setup was characterized both with an in-house toolkit\cite{groot} based on the \geant\ framework\cite{2003NIMPA.506..250G}, and a $^{152}$Eu source with an activity of 460~\ac{kBq} and a $^{60}$Co source with an activity of 60~\ac{kBq}. For a comprehensive overview, see reference~\cite{eligant-gg}. The $\gamma\gamma/\gamma$-decay data on $^{137}$Ba were collected using a $^{137}$Cs source with an activity of 336~\ac{kBq} for 49.5~days active data taking. The source had a thin circular active area with a diameter of 3~mm, encapsulated in the center of a cylindrical polymethylmetacrylate capsule with a diameter of 25~mm and 3~mm thickness, see Figure~\ref{fig:setup}.

\subsection{Energy spectra.} 
\begin{figure}
\begin{center}
 \epsfig{file=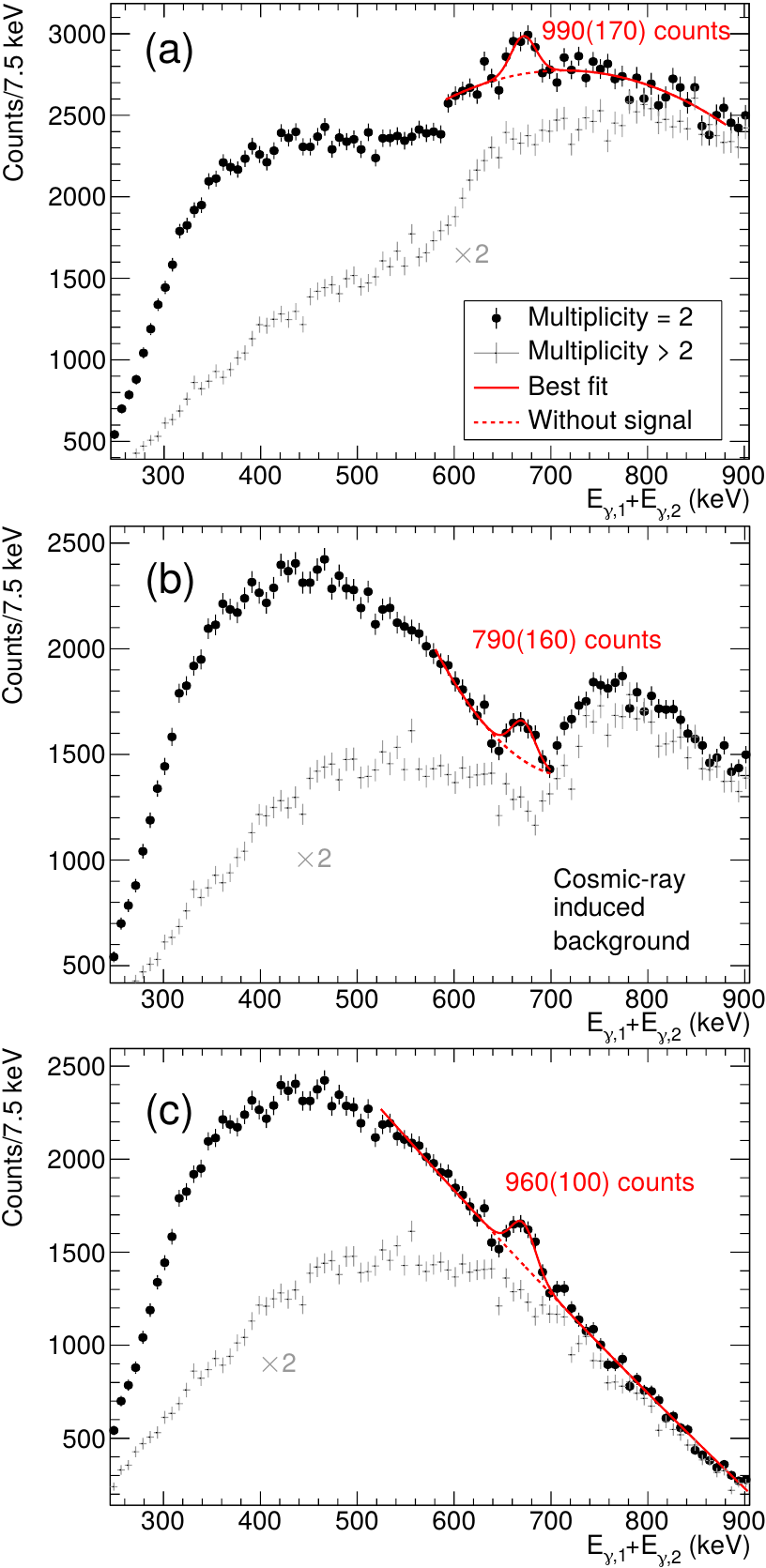,width=0.5\columnwidth}
\end{center}
\caption{{\bf{Summed double-$\gamma$ energy spectrum and data reduction.}} Black data points show the summed energy of two coincident photons detected in the CeBr$_{3}$ detectors for events with a multiplicity of two. Gray data points show the sum energy spectrum when the multiplicity is larger than two, which mainly correspond to the background induced by cosmic ray showers. We also show the fit to the data of a quadratic background as a dashed red line and the fit of the background plus a Gaussian peak as a solid red line. {\bf (a)} Raw data before any conditions. {\bf (b)} Reduced data with a condition that the energy difference between two $\gamma$ rays is $\left| E_{\gamma,1}-E_{\gamma,2}\right|<300$~keV. {\bf (c)} Final data with the additional condition that  $\left| E_{\gamma,1}-E_{\gamma,2}\right| < 960 - \left(E_{\gamma,1}+E_{\gamma,2}\right)$~keV to remove cosmic-ray induced background. The error bars represent the one standard-deviation statistical uncertainty.\label{fig:spectrum}}
\end{figure}

From the data set obtained with the $^{137}$Cs source a $(\gamma_{1},\gamma_{2})$ coincidence matrix was constructed where the $\gamma$ rays were considered coincident if the time difference between them were less than one standard deviation from the prompt time distribution, $\Delta t_{1,2} \leq 655$~ps. This condition was obtained from the coincident 444~\ac{keV} and  245~\ac{keV} $\gamma$ rays from the $2^{+}_{2}\to4^{+}_{1}\to2^{+}_{1}$ decay chain in $^{152}$Sm following the electron capture decay of $^{152}$Eu. Corrections for detector efficiencies were done on an event-by-event basis\cite{eligant-gg}. A time difference of $20 \leq \Delta t_{1,2} \leq 820$~ns was used to estimate the uncorrelated background events with two detected $\gamma$ rays and subtracted after applying an appropriate scaling factor. To remove the background contribution from electron-positron pairs produced by cosmic rays a multiplicity-two condition was assigned together with  an additional energy condition that $\left| E_{1}-E_{2}\right| < 960 - \left(E_{1}+E_{2}\right)$~\ac{keV}. The full data set, as well as the different angular groups,  were used to construct the summed energy spectra. The peaks were fitted assuming a quadratic background both with a Gaussian distribution as well as \geant\ simulated data. Both fitting methods gave consistent results. The full summed spectrum of $E_{1}+E_{2}$ with these conditions imposed is shown in Figure~\ref{fig:spectrum}.

\subsection{Branching.}

\begin{figure}
\begin{center}
 \epsfig{file=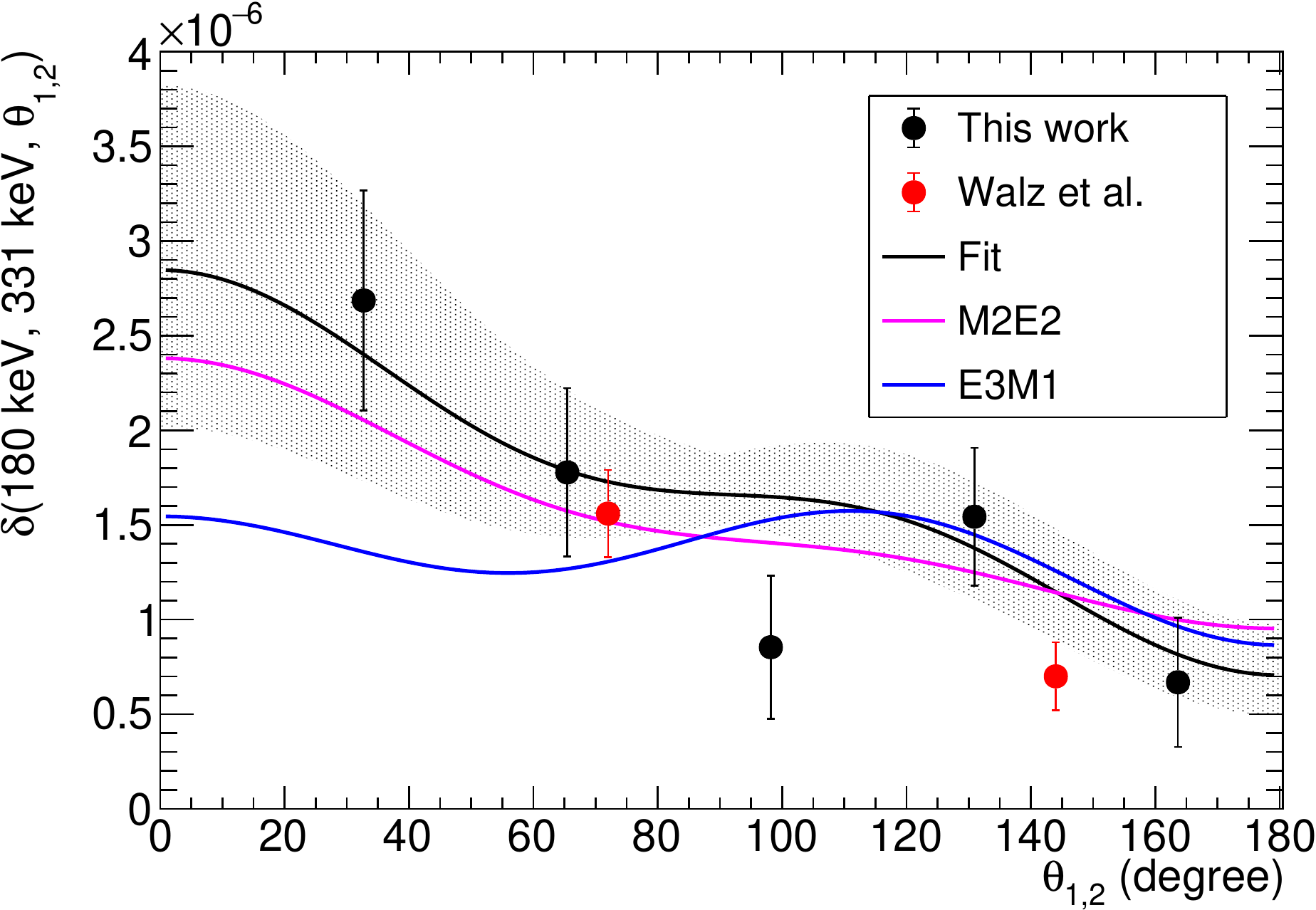,width=\columnwidth}
\end{center}
 \caption{{\bf{Angular distribution.}} The angular correlation of the two photons emitted in the double-$\gamma$ decay from this work and reference~\cite{walz_nature}, compared to the expected angular distributions of pure M2E2 and E3M1 decay. The error bars represent the one standard-deviation statistical uncertainty. \label{fig:results}}
\end{figure}

As experimental observable to evaluate the relative decay probability we use the definition of the integrated differential branching ratio\cite{walz_nature},
\begin{equation}
 \delta(E_{1},E_{2},\theta_{1,2}) = \frac{(4\pi)^{2}}{\Gamma_{\gamma}}\int_{E_{1}}^{E_{2}}\mathrm{d}\omega\left. \frac{\mathrm{d}\Gamma_{\gamma\gamma}^{5}}{\mathrm{d}\omega\mathrm{d}\Omega\mathrm{d}\Omega{'}}\right|_{\theta_{1,2}}\label{eq:delta}.
\end{equation}
In this definition $\Gamma_{\gamma}$ is the total single-gamma decay width, proportional to the size of the single-gamma peak. Given an angle, $\theta_{1,2}$, the differential decay is integrated over the frequency of the $\gamma$ ray, $\omega$. The frequency is proportional to the energy, and the integration limits are taken as the edges of the energy bin of interest. In the experimental spectrum a natural low-energy limit comes from the low-energy threshold of the detectors around 120~\ac{keV}. However, to reduce the contamination from the 511~\ac{keV} $\gamma$-rays originating from electron-positron annihilation, the integration limits $E_{1}=180$~\ac{keV} and $E_{2}=331$~\ac{keV} were chosen. The upper limit was chosen as the half of the total energy as we are not able to distinguish any relative ordering of the $\gamma$ rays. This procedure was performed for all combinations of $\theta_{1,2}$ and $\delta$ was evaluated as a function of angle. The results from this evaluation is shown in \tfig\ref{fig:results}.

 This data can be directly fitted to the generalized polarizability functions of  equation~(\ref{eq:legendre_stuff}) discussed in the methods section, using only $\alpha_{\mathrm{M2E2}}$ and $\alpha_{\mathrm{E3M1}}$ as free parameters. Other components like $\alpha_{\mathrm{E2M2}}$ or  $\alpha_{\mathrm{M3E1}}$ could in principle also contribute. However, the general polarizability functions are linearly dependent in the exchange of terms, weighted by the coefficients given by the Wigner~$6j$ symbols, and this experiment is not sensitive to this ordering. These additional components are, furthermore, expected to be small. Thus, we restrict the discussion to the $\alpha_{\mathrm{M2E2}}$ and $\alpha_{\mathrm{E3M1}}$ polarizabilities from here on.

\subsection{Energy sharing distributions.} The angular distributions themselves are not enough to completely distinguish between the contribution from the different polarizabilities. When calculating the goodness-of-fit ($\chi^{2}$), two local minima corresponding to either a large $\alpha_{\mathrm{M2E2}}$ component or a large $\alpha_{\mathrm{E3M1}}$ component appear. Instead, it is necessary to study the energy-sharing distributions between the two individual $\gamma$-rays. From equations (\ref{eq:legendre_stuff}) and (\ref{eq:polarization_functions}) in the methods section it is clear that the energy dependence of the decay for the two different cases follows $\frac{\mathrm{d}\Gamma_{\gamma\gamma}}{\mathrm{d}\omega}\propto\omega^{5}\omega{'}^{5}$ for M2E2 and as $\frac{\mathrm{d}\Gamma_{\gamma\gamma}}{\mathrm{d}\omega}\propto\omega^{3}\omega{'}^{7}$ for E3M1 with $\omega'=662-\omega$. It is clear from these relations that the energy sharing distributions are expected to have a maximum at $E_{\gamma}=E_{\gamma}'=331$~keV for the M2E2 type transitions, while an asymetric maximum is expected at $E_{\gamma}=200$~keV and $E_{\gamma}'=442$~keV for the E3M1 type transitions.

\begin{figure*}
\begin{center}
   \epsfig{file=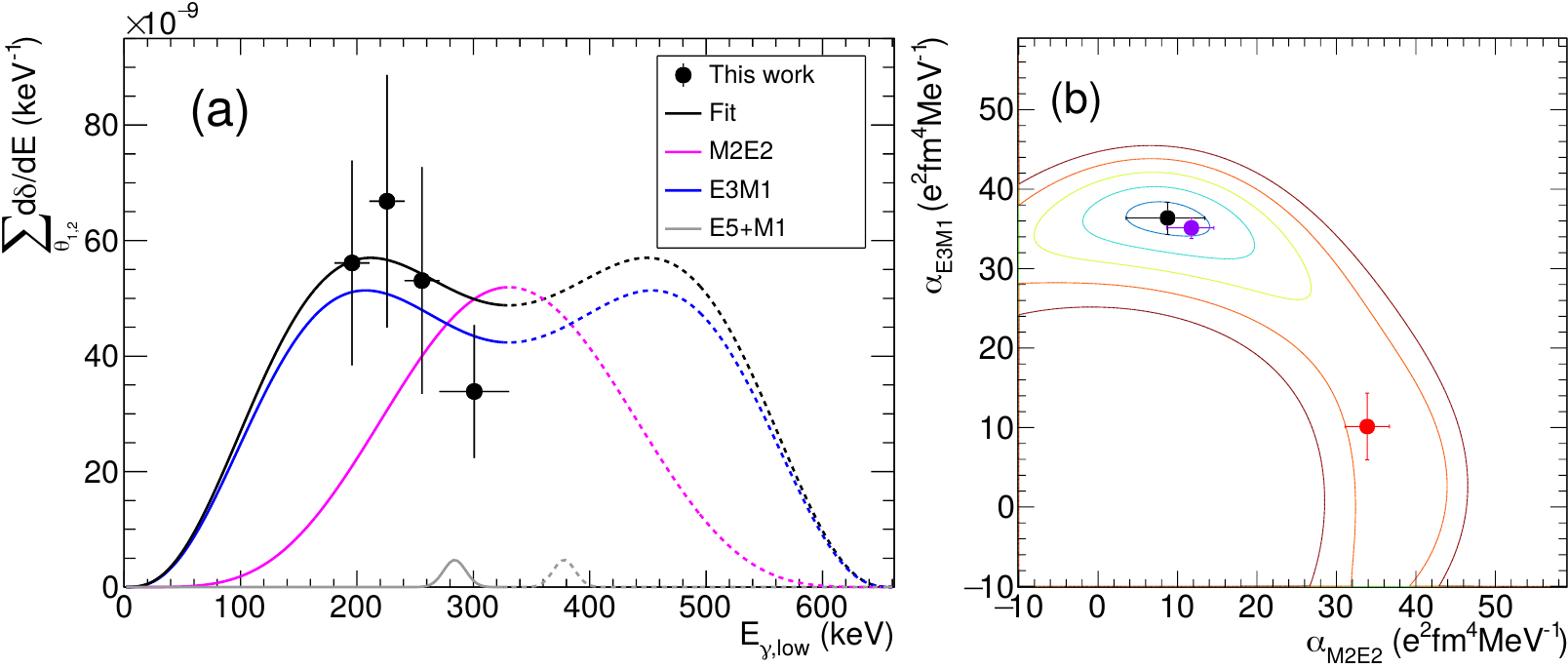,width=\textwidth}
\end{center}
\caption{{\bf{Multipole nature of the $\gamma\gamma/\gamma$ decay.}} {\bf{(a)}} Energy sharing distribution for the two photons in the double-$\gamma$ decay compared to the expected energy distributions of pure M2E2 and E3M1 decay, as well as the two-step E5+M1 decay where the measured intensity\cite{E5_decay} has been subtracted from the 300~keV data point. {The data points correspond to the sum of the differential branching ratio defined in equation~(\ref{eq:delta}) over all the available angles.} {\bf{(b)}} Two-dimensional goodness-of-fit, $\chi^{2}$, plot for the two $\alpha$ parameters with the experimental data. The contours are separated by one standard deviation. The data from the present work is shown as a black point, data from reference~\cite{walz_nature} is shown as a red point, and a fit with the two data sets combines is shown as a purple point. The error bars represent the one standard-deviation statistical uncertainty, except the error bars in $E_{\gamma,\mathrm{low}}$ that represent the width of the energy bin.\label{fig:edist}}
\end{figure*}

For this purpose, $\delta$ from equation~(\ref{eq:delta}) was evaluated in separate slices of 30~keV energy difference between the low- and high-energy limit of $E_{\gamma}$. Figure~\ref{fig:edist}a shows the results of these evaluations. A $\chi^{2}$ value was then calculated based on (\ref{eq:delta}) for different values of $\alpha_{\mathrm{M2E2}}$ and $\alpha_{\mathrm{E3M1}}$ simultaneously using the energy-integrated angular data points and the angle-summed energy data points as 
\begin{equation}
    \chi^{2} = \sum_{\mathclap{\substack{\theta_{1,2}^{i}\\E_{1}=181\\E_{2}=331}}}\frac{\delta^{2}(E_{1},E_{2},\theta_{1,2}^{i})}{\sigma^{2}_{\delta}(E_{1},E_{2},\theta_{1,2}^{i})}+\sum_{\mathclap{\substack{E_{1}=E_{\mathrm{low}}\\E_{2}=E_{\mathrm{high}}}}}\frac{\delta^{2}(E_{1},E_{2},\Sigma{\theta_{1,2}})}{\sigma^{2}_{\delta}(E_{1},E_{2},\Sigma{\theta_{1,2}})},\label{eq:chi2}
\end{equation}
where $\sigma_{\delta}$ is the statistical uncertainty in each data point, including both the signal and the subtracted background. The systematic uncertainty mainly originates from uncertainties in the intrinsic and geometric efficiencies of the setup and is expected to be on the order of a few \%, much smaller than the statistical uncertainties, and have been neglected in this expression. When including the data from Walz et al., only the energy distribution of the 72\degree\ data was included in  the second part of equation~(\ref{eq:chi2}), and the lower energy summation limit for the 144\degree\ data point was set to 206~keV in the first part of equation~(\ref{eq:chi2}).
The resulting $\chi^{2}$ surface is shown in figure~\ref{fig:edist}b. As seen here, the $\chi^{2}$ analysis from this data favours a large $\alpha_{\mathrm{E3M1}}$ component, in contradiction with both the experimental interpretation and theoretical conclusions reported in reference~\cite{walz_nature}.

\section{Discussion}

To understand these results we performed theoretical calculations of the polarization functions from equation~(\ref{eq:off_d_pol}), using the \ac{QPM}\cite{Sol76} approach. The application of the \ac{QPM} in the case of odd-mass spherical nuclei is discussed in detail in reference~\cite{Gal88}. In particular, the nuclear structure of $^{137}\mathrm{Ba}$ was studied within the framework of this model in references~\cite{Tso00,Tso06} and in reference~\cite{walz_nature}. 
In the work presented here, the calculations were built on the \ac{EDF} theory coupled with the \ac{QPM} \cite{Tso16} to obtain magnetic and electric spectral distributions. 
The model parameters of the \ac{EDF}+\ac{QPM} approach are firmly determined from  nuclear structure data or derived fully microscopically \cite{Hof98,Ton10,Gen13}.  The theoretical results are shown in table \ref{tab:alphas} and agrees with the data in terms of absolute branching strength, $\Gamma_{\gamma\gamma}/\Gamma_{\gamma}$. In addition, the \ac{EDF}+\ac{QPM} used here predicts a significantly larger $\alpha_{\mathrm{E3M1}}$ than the value reported in reference \cite{walz_nature} from the \ac{QPM}, close to our experimental observations.
However, the relative magnitude of the $\alpha_{\mathrm{M2E2}}$ and $\alpha_{\mathrm{E3M1}}$ coefficients obtained from the \ac{EDF}+\ac{QPM} theory, as well as from reference \cite{walz_nature}, are different than the experimental results obtained in this work. In particular, the present measurement indicates that the $\alpha_{\mathrm{M2E2}}$ coefficient is significantly smaller than previously reported, and at this level of complexity \ac{EDF}+\ac{QPM} it is not able to account for the apparent discrepancy with the experimental data. 

To understand the origin of this discrepancy, the properties of the dominant, low-lying, states were investigated from another perspective using the state-of-the-art nuclear \ac{MCSM}\cite{mcsm1,mcsm2}. These calculations were used to extract information from the three lowest-energy $J^{\pi}=7/2^{+}$ states, the five lowest-energy $J^{\pi}=5/2^{+}$ states, as well as the ground $J^{\pi}=3/2^{+}$ and isomeric $J^{\pi}=11/2^{-}$ states.
The neutron component in the \ac{MCSM} wave function of the isomeric $J^{\pi}=11/2^{-}$ state is dominated by 
a single neutron hole in $\nu h_{11/2}$. The $J^{\pi}=7/2^{+}$ state is different, however, with most of the neutron hole occupation is in $\nu d_{3/2}$, 
coupled to a $2^{+}$ state of six valence protons. 
The $\nu g_{7/2}$ orbital itself is almost full. 
This is in contrast with the \ac{EDF}+\ac{QPM} results where the $2^{+}\otimes \nu d_{3/2}$ contribution is 38.7\% and the $\nu g_{7/2}$ single-particle component is 51.3\%. 
Thus, the odd-neutron contribution to the M2 transition rate in the \ac{MCSM} would require a highly hindered transition between $\nu h_{11/2}$ and $\nu d_{3/2}$, or by utilising a minor $\nu g_{7/2}$ vacancy. This, gives rise to a strongly hindered M2 transition within the \ac{MCSM}, with a reduced transition probability, $B(\mathrm{M2})=13.5\times 10^{-3}$~$\mu^{2}$fm$^{2}$, three orders of magnitudes less than predicted by the \ac{EDF}+\ac{QPM} model where $B(\mathrm{M2})=14.9$~$\mu^{2}$fm$^{2}$. This can explain the observed suppression of $\alpha_{\mathrm{E2M2}}$. It is interesting to note that with increasing excitation energy, the \ac{MCSM} predict a smooth change in orbital occupation from $\nu d_{3/2}$ to $\nu d_{5/2}$, constructively adding to the M2 transition strength for all the calculated $7/2^{+}$ transitions in contrast to the \ac{EDF}+\ac{QPM} where all higher-lying states act destructively. Table~\ref{tab:matrixelements} lists the contributing low-lying matrix elements discussed here.

\begin{table*}[h!]
\centering
\begin{tabular}{|p{0.8\textwidth}|}
\hline
\caption{\bf{Calculated  matrix elements.\label{tab:matrixelements}}}
 \begin{tabular*}{0.8\textwidth}{@{\extracolsep{\fill}}lcclcc}
\hline
 Matrix element & EDF+QPM & MCSM & Matrix element & EDF+QPM & MCSM \\
   & $\mathrm{e}\cdot\mathrm{fm}^{L}$ & $\mathrm{e}\cdot\mathrm{fm}^{L}$ &    & $\mathrm{e}\cdot\mathrm{fm}^{L}$ & $\mathrm{e}\cdot\mathrm{fm}^{L}$ \\
\hline
$\langle 3/2^{+}_{1} \| \mathrm{M1} \| 5/2^{+}_{1} \rangle$ & {\bf{-0.11}} & {\bf{-0.139}}  & $\langle 5/2^{+}_{1} \| \mathrm{E3} \| 11/2^{-}_{1} \rangle$ & {\bf{-168}} & {\bf{57.2}}\\
$\langle 3/2^{+}_{1} \| \mathrm{M1} \| 5/2^{+}_{2} \rangle$ & -0.03 & {\bf{0.172}} & $\langle 5/2^{+}_{2} \| \mathrm{E3} \| 11/2^{-}_{1} \rangle$ & -57.3 & {\bf{-81.9}}\\
$\langle 3/2^{+}_{1} \| \mathrm{M1} \| 5/2^{+}_{3} \rangle$ & -0.04 & {\bf{0.183}} & $\langle 5/2^{+}_{3} \| \mathrm{E3} \| 11/2^{-}_{1} \rangle$ & -90.2 & {\bf{-128}}\\
$\langle 3/2^{+}_{1} \| \mathrm{E2} \| 7/2^{+}_{1} \rangle$ & {\bf{63.9}} & {\bf{39.0}} & $\langle 7/2^{+}_{1} \| \mathrm{M2} \| 11/2^{-}_{1} \rangle$ & {\bf{1.14}} & {\bf{-0.0518}}\\
$\langle 3/2^{+}_{1} \| \mathrm{E2} \| 7/2^{+}_{2} \rangle$ & {\bf{-46.4}} & 4.99 &$\langle 7/2^{+}_{2} \| \mathrm{M2} \| 11/2^{-}_{1} \rangle$ & {\bf{0.76}} & -0.112 \\
\hline
 \end{tabular*} 
\footnotesize{Transition matrix elements of the lowest-energy transitions, in each model, calculated using the EDF+QPM and MCSM models for the  states that contribute to the double-$\gamma$ decay in $^{137}\mathrm{Ba}$. The EDF+QPM values for the magnetic transitions correspond to ${g_{\mathrm{s}}^{\mathrm{eff}}=0.6g_\mathrm{s}^{\mathrm{bare}}}$ while the MCSM values correspond to ${g_{\mathrm{s}}^{\mathrm{eff}}=g_\mathrm{s}^{\mathrm{bare}}}$. The states that dominates the decay in each model have been highlighted with {\bf{bold}} font.}
\hrule
\end{tabular}
\end{table*}

Regarding the $\alpha_{\mathrm{E3M1}}$ component of the decay, 
the main components obtained from the \ac{EDF}+\ac{QPM} calculations are from the coupling of the single-particle mode with the surface vibrations of the even-even core. 
As a consequence, due to the exchange of the collective $3^{-}_1$ octupole phonon, we obtain a rather strong E3 transition, consistent with our experimental observations. 
For these states the \ac{EDF}+\ac{QPM} and the \ac{MCSM} give a consistent picture with a constructive addition to the strength for each successive state among the first three excited states with the main difference that in the \ac{EDF}+\ac{QPM}, the main contribution comes from the $5/2^{+}_{1}$ state while the \ac{MCSM} predicts that the $5/2^{+}_{2,3}$ states are dominating.


\begin{methods}

{\subsection{Experimental setup.} The set-up consisted of eleven 3''$\times$3'' CeBr$_{3}$ detectors coupled with Hamamatsu R6233 photomultiplier tubes and built-in voltage dividers. The high voltage for the photomultiplier tubes were provided by a CAEN SY4527 power supply. The signals were read out using one CAEN V1730 digitizer operating with a 14-bit resolution at a 500~MS/s sampling rate and a dynamic range of 0.5~V$_{\mathrm{pp}}$, running PSD firmware. The digitizers were controlled using the \ac{MIDAS} software and triggered individually. Each event consisted of the energy, the time-stamp, and the the digitized voltage pulse from the detector. The sub-nanosecond time information was obtained from the value of the time-stamp corrected by a digital interpolation of the sampling points in the recorded pulse, at half of the maximum value of the pulse and interpolated using a quadratic polynomial.}

\subsection{Polarization functions.} To obtain the nuclear polarizabilites, $\alpha_{S'L'SL}$, from the differential decay probability we follow the theoretical treatment in references \cite{walz_nature,gg_16O}. Here the differential decay probability can be expressed in terms of generalized polarization functions, $P_{J}'(S,L,S',L')$, and Legendre polynomials, $P_{l}(\cos \theta)$, as 
\begin{equation}
 \frac{\mathrm{d}^{5}\Gamma_{\gamma\gamma}}{\mathrm{d}\omega\Omega\Omega'}=\frac{\omega\omega'}{96\pi}\sum P_{J}'(S{'}_{1}L{'}_{1}S_{1}L_{1})P_{J}'(S{'}_{2}L{'}_{2}S_{2}L_{2})\sum a_{l}^{\xi} P_{l}(\cos\theta),\label{eq:legendre_stuff}
\end{equation}
where the generalized polarization functions are defined as
 \begin{equation}
 \begin{split}
  P_{J}'(S'L'SL) = &(-1)^{S+S'}2\pi(-1)^{I_{i}+I_{f}}\omega^{L}\omega{'}^{L'}\cdot\\&\sqrt{\frac{L+1}{L}}\sqrt{\frac{L'+1}{L'}}\frac{\sqrt{2L+1}\sqrt{2L'+1}}{(2L+1)!!(2L'+1)!!}\cdot\\&\left( \left\{ \begin{matrix}
   L & L' & J \\
   I_{f} & I_{i} & I
  \end{matrix}
\right\} \alpha_{S'L'SL}+(-1)^{S+S'}\left\{ \begin{matrix}
   L' & L & J \\
   I_{f} & I_{i} & I
  \end{matrix}
\right\}\alpha_{SLS'L'}\right).
 \end{split}\label{eq:polarization_functions}
 \end{equation}
The sums in equation~(\ref{eq:legendre_stuff}) run over all the permutations of electric, $S=0$, and magnetic, $S=1$, combinations with multipolarity, $L$, allowed in the decay, and over all Legendre polynomials with non-zero coefficients. The general polarizability functions in equation~(\ref{eq:polarization_functions}) consist of a linear combination of the off-diagonal polarizabilities of the nucleus weighted by coefficients determined by the corresponding angular momentum algebra of the decay.
\subsection{The quasiparticle-phonon model.}
The \ac{QPM} Hamiltonian includes mean field, pairing interaction and separable multipole and spin-multipole interactions \cite{Sol76}. The mean field for protons and neutrons is defined as a Woods-Saxon potential with parameter sets derived self-consistently from a fully microscopic \ac{HFB} calculations described in \cite{Tso08,Tso16}. The method assures a good description of nuclear ground-state properties by enforcing that measured separation energies and nuclear radii are reproduced as close as possible~\cite{Tso08}. 
The pairing and residual interaction parameters are fitted to reproduce the odd-even mass differences of neighbouring nuclei as well as the experimental values of the excitation energies and reduced transition probabilities of low-lying collective and non-collective states in the even-even core nucleus\cite{Sol76}. Of particular importance in these studies is the determination of the isovector spin-dipole coupling constant which is extracted from comparison to data from \cite{Ton10} and fully self-consistent \ac{QRPA} calculations using the microscopic \ac{EDF} of \cite{Hof98}.
Single-particle (s.p.) energies of the lowest-lying excited states in $^{137}\mathrm{Ba}$ are fine-tuned to experimental values to achieve the highest accuracy in the description of the experimental data. We point out that the s.p. energies problem is not a matter of the interaction parameters but originate in the quasiparticle spectrum, which indicates the necessity to go beyond the static mean-field formalism\cite{Len19,Tso19}.
 
 In the \ac{QPM} the wave functions of the excited states of an even-odd nucleus are constructed from a combination of quasiparticles originating from the single-particle orbitals and excitation phonons that are constructed from the excited states in the neighboring even-even core nucleus :
\begin{equation}
\begin{aligned}
&\Psi_{\nu} (JM) ={}C^{\nu}_J\left\{ \alpha^+_{JM} + \sum_{\lambda\mu i} D_j^{\lambda i}(J\nu) [\alpha_{jm}^+ Q^{+}_{\lambda\mu i}]_{JM}
\right\}\Psi_0
\end{aligned}
\label{wf}
\end{equation}
The notation $\alpha_{jm}^+$ is the quasiparticle creation operator with shell quantum numbers $j \equiv [(n,l, j)]$ and projection ${m}$; $Q^{+}_{\lambda\mu i}$ denotes the phonon creation operator with the angular momentum $\lambda$, projection ${m}$ and \ac{QRPA} root number ${i}$; $\Psi_0$ is the ground state
of the neighboring even-even nucleus and $\nu$ stands for the number within a sequence of states of given angular momentum $J^{\pi}$ and projection ${M}$. The coefficients $C^{\nu}_J$ and $D_j^{\lambda i}(J\nu)$ are the quasiparticle and 'quasiparticle $\otimes$ phonon' amplitudes for the $\nu$ state.
The coefficients of the wave function (\ref{wf}) and the energy of the excited states are found by diagonalisation of the model Hamiltonian within the approximation of the commutator linearization \cite{Sol76,Gal88}. The components  $[\alpha_{jm}^+ Q^{+}_{\lambda\mu i}]_{JM}$ of the wave function (\ref{wf}) may violate the Pauli principle. The exact commutation relations between quasiparticle and phonon operators are used to solve this problem. 
The properties of the phonons are determined by solving \ac{QRPA} equations from Refs.\cite{Sol76,Gal88}. The model basis includes one-phonon states with spin and parity $J^{\pi}=1^{\pm},2^{\pm},3^{\pm},4^{\pm},5^{\pm}$ and excitation energies up to ${E_x=20~MeV}$. The calculations of the $\alpha$-coefficients of the double-$\gamma$ decay probability of $^{137}$Ba include all low-energy excited states with spin and parity $J^{\pi}=1/2^{\pm}, 3/2^{\pm},5/2^{\pm}, 7/2^{\pm}, 9/2^{\pm}$ and excitation energies up to ${E_x=10~MeV}$.

In the case of the E1 transitions, we have used effective charges ${e_\mathrm{p}^\mathrm{eff}=(N/A)e}$ (for protons) and ${e_\mathrm{n}^\mathrm{eff}=-(Z/A)e}$ (for neutrons) to separate the center of mass motion and 'bare' values for $\mathrm{E2}$ and $\mathrm{E3}$ transitions ${e_\mathrm{p}=e}$ (for protons) and ${e_\mathrm{n}=0}$ (for neutrons), where ${e}$ is the electron charge.
 Following previous \ac{QPM} calculations \cite{Gen13}, the magnetic transitions are calculated with a quenched effective spin-magnetic factor ${g_{\mathrm{s}}^{\mathrm{eff}}}$. The influence of the ${g_{\mathrm{s}}^{\mathrm{eff}}}$ parameter on the experimental observables related to electromagnetic transitions of lowest-lying states and double-$\gamma$ decay probability coefficients was investigated by carrying out 
\ac{EDF}+\ac{QPM} calculations for several choices of this parameter between 0.6 and 1 of the value of the 'bare' spin-magnetic moment, ${g_\mathrm{s}^{\mathrm{bare}}}$.
 The theoretical observations indicate that the values ${g_{\mathrm{s}}^{\mathrm{eff}}=0.6-0.7g_\mathrm{s}^{\mathrm{bare}}}$ which are in agreement with our previous findings\cite{Tso00,Tso06,Gen13} reproduce quite well the experimental data on M1 and M2 transition strengths and the angular distribution of the two photons of the double-$\gamma$ decay.
 
\subsection{Monte Carlo shell model.}
In the \ac{MCSM}, the approximated wave functions, $\left|\Psi_{N_{b}}\right>$, are obtained as a superposition of spin ($I$) and parity ($\pi$) projected Slater determinant basis states, $\left|\phi_{n}\right>$, 
\begin{equation}
    \left|\Psi_{N_{b}}\right>=\sum_{n=1}^{N_{b}}\sum_{K=-I}^{I}f_{n,K}^{N_{b}}P_{MK}^{I\pi}\left|\phi_{n}\right>,
\end{equation}
where ${N_{b}}$ is the number of basis states, $P_{MK}^{I\pi}$ is the spin-parity projection operator, and the $f_{n,K}^{N_{b}}$ coefficients are obtained from diagonalizing the Hamiltonian. The set of basis states are selected by Monte Carlo methods and iteratively refined to minimize the ground state energy. The model space for these calculations included the $1g_{9/2}$, $1g_{7/2}$, $2d_{5/2}$, $2d_{3/2}$, and $3s_{1/2}$ even-parity orbitals, as well as the $1h_{11/2}$, $2f_{7/2}$, and $3p_{3/2}$ odd-parity orbitals. The two-body matrix elements were obtained from the JUN45 and SNBG3 data sets\cite{jun45,snbg3}, and the $V_{\mathrm{MU}}$ interaction\cite{vmu}. To obtain the transition matrix elements effective proton and neutron charges ${e_\mathrm{p}=1.25}$ and ${e_\mathrm{n}=0.75}$, and gyromagnetic factors ${g_{\ell,\mathrm{p}}=1}$, ${g_{\ell,\mathrm{n}}=0}$, ${g_\mathrm{s,p}=5.586}$, and ${g_\mathrm{s,n}=-3.826}$ was used.
The calculations followed the procedure for the tin isotope chain closely\cite{PhysRevLett.121.062501}. Said reference and references within contains a detailed description of the procedure.
\end{methods}


\begin{addendum}
 \item[Author contributions] P.-A.S., L.C., E.A., D.L.B., C.M., and A.P. designed the experimental setup. P.-A.S., L.C., E.A., G.L.G., D.L., D.N., and T.P. collected the data. L.C., and D.N. wrote the software for data conversion. P.-A.S. wrote the software for data analysis and analysed the data. T.O., N.T., Y.T., and H.L. performed the theoretical calculations. P.-A.S. and D.L. performed the \geant\ simulations. P.-A.S., L.C., T.O., N.T., D.L.B., and N.P. discussed the interpretation of the experimental and theoretical results. P-A.S. and N.T. prepared the manuscript draft, and all authors read and contributed to the discussion of the final manuscript.
 \item[Competing Interests] {The authors declare that they have no competing interests.}
 \item[Data availability] Raw data were obtained at the Extreme Light Infrastructure -- Nuclear Physics facility, Romania. All the data used to support the findings of this study are available from the authors upon reasonable request. {The final data points can be obtained from \url{http://dx.doi.org/10.17632/skhmjshxdj}}. 
 \item[Code availability] Sorting codes were developed at the Extreme Light Infrastructure -- Nuclear Physics facility, Romania. All the codes for the experimental data used in this study are available from the authors upon reasonable request.
 \item We acknowledge A. Imreh from ELI-NP for the CAD drawings of the detector system used for the \geant\ simulations and figure~1. The authors P.-A.S. and N.T. acknowledge the support of the Romanian Ministry of Research and Innovation under research contract 10N~/~PN~19~06~01~05. L.C., E.A. D.L.B., G.L.G., D.L., C.M., D.N., A.P., and T.P. would like to acknowledge the support from the Extreme Light Infrastructure Nuclear Physics (ELI-NP) Phase II, a project co-financed by the Romanian Government and the European Union through the European Regional Development Fund - the Competitiveness Operational Programme (1/07.07.2016, COP, ID 1334). N.P. thanks the DFG for support under grant No. SFB~1245 and the state of Hesse for support of the ``Nuclear Photonics'' project within the LOEWE initiative. The MCSM calculations were performed on the K computer at RIKEN AICS, Project Number hp190160, and Oakforest-PACS operated by JCAHPC. This work was supported in part by the Post-K Computer Priority Issue ``Elucidation of the Fundamental Laws and Evolution of the Universe'' from MEXT and JICFuS, and the Multidisciplinary Cooperative Research Program at the CCS, University of Tsukuba. We also would like to thank Dr.~D.~Gambacurta at ELI-NP for many valuable discussions.  
 \item[Correspondence] Correspondence and requests for materials should be addressed to P.-A.S.~(email: par.anders@eli-np.ro).
\end{addendum}




\acrodef{AGATA}{Advanced GAmma Tracking Array}
\acrodef{CFD}{constant-fraction discriminator}
\acrodef{DAQ}{data acquisition system}
\acrodef{ELI-NP}{Extreme Light Infrastructure -- Nuclear Physics}
\acrodef{ELI}{Extreme Light Infrastructure}
\acrodef{ELIGANT}{ELI Gamma Above Neutron Threshold}
\acrodef{ELIGANT-GG}{ELIGANT Gamma Gamma}
\acrodef{ELIGANT-GN}{ELIGANT Gamma Neutron}
\acrodef{ELIGANT-TN}{ELIGANT Thermal Neutron}
\acrodef{EOS}{equation of state}
\acrodef{FWHM}{full-width at half-maximum}
\acrodef{GDR}{giant dipole resonance}
\acrodef{GT}{Gamow-Teller}
\acrodef{HPGe}{high-purity germanium}
\acrodef{ISOL}{Isotope Separation On-Line}
\acrodef{kBq}{kilo Becquerel}
\acrodef{keV}{kilo electron-volt}
\acrodef{MCSM}{Monte Carlo shell model}
\acrodef{MIDAS}{Multi Instance Data Acquisition System}
\acrodef{PDR}{pygmy dipole resonance}
\acrodef{QPM}{quasiparticle-phonon model}
\acrodef{QRPA}{quasiparticle random phase approximation}
\acrodef{TOF}{time-of-flight}
\acrodef{HFB}{Hartree-Fock-Bogoljubov}
\acrodef{EDF}{energy-density-functional}


\end{document}